\documentstyle[epsfig]{mn}

\title[Ionizing radiation in {\sc sph}]{Ionizing radiation in Smoothed
Particle Hydrodynamics}

\author[O.~Kessel-Deynet, A.~Burkert]{O.~Kessel-Deynet, A.~Burkert\\
	Max-Planck-Institut f\"ur Astronomie, K\"onigstuhl 17, D-69117
	Heidelberg, Germany}

\begin{document}

\maketitle
\begin{abstract}
A new method for the inclusion of ionizing radiation
from uniform radiation fields into 3D Smoothed Particle Hydrodynamics
({\sc sphi})
simulations is presented. We calculate the optical depth for the Lyman continuum
radiation from the source towards the {\sc sphi} particles by ray-tracing
integration. The time-dependent ionization rate equation is then solved locally for the
particles within the ionizing radiation field. Using test calculations, we explore
the numerical behaviour of the code with respect to the implementation
of the time-dependent ionization rate equation. We also test the
coupling of the heating caused by the ionization to the hydrodynamical
part of the {\sc sphi} code. 
\end{abstract}

\section{Introduction}

Smoothed Particle Hydrodynamics ({\sc sph}) has become a numerical method
widely used for addressing problems related to fluid flows in
astrophysics. Due to its Lagrangian nature it is especially well suited
for applications involving variations by many orders of magnitude in density. Examples for this type of applications are simulations of the collapse
of molecular clouds and the formation of a stellar cluster, as
performed by Klessen, Burkert \& Bate~\shortcite{klebuba}. A comparison between grid based methods
and {\sc sph} was performed by Burkert, Bate \& Bodenheimer~\shortcite{bbb} and Bate \&
Burkert~\shortcite{bateburk}. They applied both
methods to the numerically demanding problem of gravitational collapse and
fragmentation of a slightly perturbed rotating cloud core with an
$r^{-2}$ density profile. Both methods yielded the same qualitative
results.
%% Other comparisons between different numerical methods,
%%a tree sph code~\cite{HernKatz}, a P$^3$M sph scheme~\cite{Evrard} and
%%several Eulerian schemes~\cite{Ryu,Brian,Cen}, was carried out by Kang et
%%al.~\shortcite{Kang}. They found satisfactory convergence of all
%%methods to the same results with increasing resolution. 
Bate~\shortcite{bate2} performed the first
calculation which followed the collapse of a molecular cloud core in
3 dimensions down to a protostellar object in hydrodynamical equilibrium, thus spanning
17 (!) orders of magnitude in density. Other applications include
accretion processes in massive circumbinary disks \cite{boba1,babo1},
the collapse of cloud cores induced by shock waves \cite{vanhala} or
colliding clumps \cite{bhattal}, the precession of accretion disks in binary systems
\cite{larwood}, the dynamical behaviour of massive protostellar disks
\cite{nelson} or the formation of large scale structure and galaxies in
the early universe \cite{steinmetz}.
 
A variety of physical processes are at work in the interstellar medium, like magnetic
fields, radiation or thermal conductivity, necessitating their
inclusion into numerical codes. This has already
been achieved to a large extent in grid based methods like the magneto-hydrodynamics
codes {\sc zeus} \cite{stono} or {\sc nirvana} \cite{ziegler},
or codes including effects of IR and UV radiation \cite{yorka,sonnha,richling}.

In contrast, the addition of physical processes to {\sc sph} codes is just
at its beginnings. Extensions
achieved so far are sophisticated equations of state (e.g. Vanhala
et al.~1998) and self-gravity. Some efforts were made to make {\sc sph} faster and
more accurate. The introduction of {\sc tree} algorithms \cite{barnes,press,benz} and the use of GRAvity PipE
({\sc grape}), a hardware device for fast computation of the gravitational N-body
forces \cite{umemura,steinmetz}, helped reducing the
numerical effort for the gravitational force calculation and the
determination of the nearest neighbours for each
particle. Inutsuka~\shortcite{inutsuka} presented a Godunov-like solver for
the Eulerian equations in {\sc sph} thus enhancing the numerical treatment of
shocks. The introduction of gravitational periodic boundaries~\cite{hernqu,klessen} allows the treatment of fragmentation and
turbulence in molecular clouds without global collapse. The timestep
problem which arises during isothermal collapse calculations at high
densities is circumvented by the formation of sink particles, which
substitute the innermost parts of the collapsing clump by one particle
and accumulate the infalling mass and momenta \cite{babopri}.

The strength of {\sc sph} lies in its Lagrangian
nature, which makes it especially attractive for problems involving
gravitational collapse and star formation. Applications like e.g. by Klessen
et al.~\shortcite{klessen}, which deal with the collapse
and fragmentation of molecular clouds, neglect the feedback processes
of newly born stars which act on their parental cloud through stellar
winds, outflows and ionization. This simplification may be justified
as long as the simulations deal with collapse on timescales smaller
than $\approx$ 1 Myr, on which single and binary stars or T Tau-like
clusters are formed~\cite{efremov}. The case is different for larger
timescales, on which OB subgroups and associations are
formed. Neglecting feedback in these cases can lead  to
unphysical results, like a star formation efficiency of 100 per cent,
since in the purely isothermal case all material will sooner or later
be accreted onto the evolving protostellar cores. This is in strong
contradiction to observations, which estimate a global star formation
efficiency for ordinary molecular clouds of order 10 per cent
\cite{pppII}. Another possible effect of feedback is the induction of
star formation due to the compression of cloud material by shock waves
and ionization fronts. 

In this paper we discuss the implementation of the effects of
ionizing UV radiation by massive stars into {\sc sph} calculations as a first
step in order to perform collapse calculations on scales where
OB-stars are formed in a more realistic
way. This will in future applications allow us to assess questions
like: How does the process of ionization by massive stars change the
stellar initial mass function? What are the implications for the star formation
efficiency? Can star formation be induced by
ionization, and if yes, what are the time scales and the parameter
space, for which induced star formation can be expected? These
questions will be discussed in subsequent papers.

%%Some of the questions above were already adressed in the
%%past. Bertoldi~\shortcite{bertoldi2} and Bertoldi \&
%%McKee~\shortcite{bertoldi1} developed an analytic theory for     

\section{Physical problem}

We incorporate the effects of ionizing radiation from hot stellar
photospheres into {\sc sph} by dividing the problem into three major substeps:
\begin{enumerate}
\item calculation of the UV radiation field by solving the time-independent, non-relativistic equation of radiative transfer,
\item determination of the ionization and recombination rates from the
local radiation field, density and ionization fraction,
\item advancing the ionization state of the particles in time by
solving the time-dependent ionization rate equation.
\end{enumerate}

\subsection{Calculation of the UV radiation field}

Given a planar infall of ionizing photons from a distant source onto
the border of the volume of interest with a flux $J_0$ Lyman
continuum photons per time and square area, the resulting
photon flux inside this volume is given by
\[
  J(s)=J_0 \cdot \exp \left( - \tau \left( s \right) \right),
\]
where $\tau(s)$ is the optical depth for the ionizing photons along
the line of sight parallel to the infall direction of the photons, and
$s$ is the distance from the border of the integration volume along
the line of sight: 

\begin{equation} \label{eq:LOS}
  \tau(s) = \int_{0}^{s} \left[ \bar{\kappa} \left( s^{\prime} \right) +
  \kappa_{\rm d} \left( s \right) \right] {\rm d} s^{\prime}.
\end{equation}
 
%%where $\tau$ is the optical depth for the ionizing photons along the
%%LOS:
%%\begin{equation} \label{eq:LOS}
%%  \tau^{({\rm i})} = \int_{{\bf r}^{({\rm i})}}^{\bf r} \bar{\kappa}
%%  \left( s \right) {\rm d}s.
%%\end{equation} 
We neglect the effect of `photon hardening', i.e. the stronger
absorption of weaker photons, and use an `effective' absorption
coefficient $\bar{\kappa}$,
the mean of $\kappa_{\nu}$ over frequency, weighted by the spectrum of
the source $S_{\nu}$:
\begin{equation} \label{eq:crossect}
  \bar{\kappa} = n_{\rm H} \cdot \bar{\sigma} = n_{\rm H} \cdot
  \frac{\int S_{\nu}^{({\rm i})} \sigma_{\nu} {\rm
  d} \nu}{S_{\rm tot}^{({\rm i})}},
\end{equation}
where $\sigma_{\nu}$ denotes the ionization cross section of hydrogen
in the ground state and $n_{\rm H}$ the particle density of the H atoms.

The role of dust in H{\sc ii} regions and its effect on ionizing
radiation is still very uncertain~\cite{necklace}. If dust is present,
it will partially absorb UV photons, heat up and reemit the energy in
the IR regime. Its first order effect can be included easily under the
assumption of a homogeneous distribution of the dust in the H{\sc ii}
region. The corresponding contribution to the optical depth can be
incorporated by adding the dust absorption coefficient at the Lyman
border $\kappa_{\rm d}$ to the absorption coefficient in
Eq.~\ref{eq:LOS}. $\kappa_{\rm d}$ depends on the dust model used and
is regularily determined using Mie theory for grains with given
distributions in size and shape. In this paper, we set $\kappa_{\rm
d}$ to zero throughout.  

%%The flux of ionizing photons in the case of point sources as the only
%%emitters is given by the superposition:
%%\begin{equation}
%%  {\bf J} = \sum \frac{J^{({\rm i})}}{|{\bf r}^{({\rm i})}|} \cdot {\bf r}^{({\rm
%%  i})}.
%%\end{equation}
We also neglect the diffuse field of Lyman continuum photons, which
are being produced by recombinations of electrons into the ground
level and which themselves possess sufficient energy for ionizing
other H atoms. A thorough treatment of this radiation can only be
achieved by detailed radiation transfer calculations as proposed
e.g. by Yorke \& Kaisig~\shortcite{yorka}. Instead we use the
assumption of the validity of the `on the spot' approximation as
follows: due to the fact that the spectrum of the Lyman
recombination photons as well as the the ionization cross section is
strongly peaked at the Lyman border, a small
amount of H atoms in the ionized region is sufficient to make the
medium optically thick for the Lyman recombination photons. This leads
to the absorption of these photons in the ultimate vicinity of their
creation sites. As the creation of one photon is related to
the creation of one H atom, its absorption leads to the
destruction of one H atom. Thus the net effect of these photons on the
local ionization structure is zero.

This assumption breaks down in regions next to OB stars, where due to
the high UV flux the density of H atoms is not sufficient to make the
medium optically thick to Lyman continuum photons. Next to ionization
fronts, where the density of H atoms is much higher, the `on the spot'
assumption is nevertheless a good approximation. On further details
refer to Yorke~\shortcite{yorkebook}.

\subsection{Ionization and recombination rates}

The ionization rate in the medium is given by the sinks of the UV
radiation field, since every ionization leads to the absorption of one
UV photon:
\begin{equation} \label{eq:ionrate}
  {\cal I} = n_{\rm H} \bar{\sigma} J = -\nabla \cdot {\bf J},
\end{equation}
where ${\bf J} = J \hat{{\bf e}}_s$ is the flux vector in the
direction $\hat{{\bf e}}_s$ of the line of sight.

The recombination rate can be estimated as :
\begin{equation} \label{eq:recomb}
  {\cal R}=n_{\rm e}^2 \alpha_{\rm B}=n^2x^2 \alpha_{\rm B},
\end{equation}
with $n$ being the particle density of H atoms and protons together,
$n_{\rm e}$ the particle density of free electrons, $x=n_{\rm e}/n$
the ionization fraction and $\alpha_{\rm B}$ the effective
recombination coefficient under assumption of validity of the `on the spot'
approximation. The recombination coefficient $\alpha$ is given as the
sum over the individual recombination coefficients $\alpha_n$, where
the electron ends up in the atomic level $n$:
\begin{equation} \label{eq:alpha}
  \alpha = \sum_{n} \alpha_n.
\end{equation}
Under the assumption of the `on the spot' approximation recombinations
into the ground level do not lead to any net effect and thus
$\alpha_1$ can be neglected in Eq.~\ref{eq:alpha}. The resulting net
recombination rate which is used in Eq.~\ref{eq:recomb} is commonly
called $\alpha_{\rm B}$ after the nomenclature introduced by Baker \&
Menzel~\shortcite{baker}:
\[
   \alpha_{\rm B} = \sum_{n=2}^\infty \alpha_n.
\]

\subsection{Ionization rate equation}

Knowing the ionization and recombination rates, ${\cal I}$ and ${\cal
R}$, the ionization fraction can be calculated from the ionization rate
equation. The time dependency of the ionization fraction in the frame
comoving with the corresponding particle, i.e. its Lagrangian
formulation, is given by:
\begin{equation} \label{eq:ioniztime}
  \frac{{\rm d} n_{\rm e}}{{\rm d} t} = {\cal I}-{\cal R}.
\end{equation}

\subsection{Modeling the source}

Since the spectral distribution of the UV radiation emitted by
the photospheres of intermediate to high mass stars is very uncertain,
we assume a black radiator with an effective temperature $T_{\star}$.

\section{Numerical treatment}

We developed two different methods for the numerical treatment of time
dependent ionization in the {\sc sph} calculations. Both have in common the
method of finding paths from the ionizing source to the particles,
along which the optical depth for the Lyman continuum photons can be
calculated. They differ in the way the ionization rate is determined
given the radiation field. Method A uses the {\sc sph} formalism to calculate
the divergence of the radiation field in Eq.~\ref{eq:ionrate}. In method B
we adopt a different approach also used in grid methods, where we
derive the ionization rate from the difference in the numbers of
photons entering and leaving a particle.
\begin{figure*}
  \epsfig{file=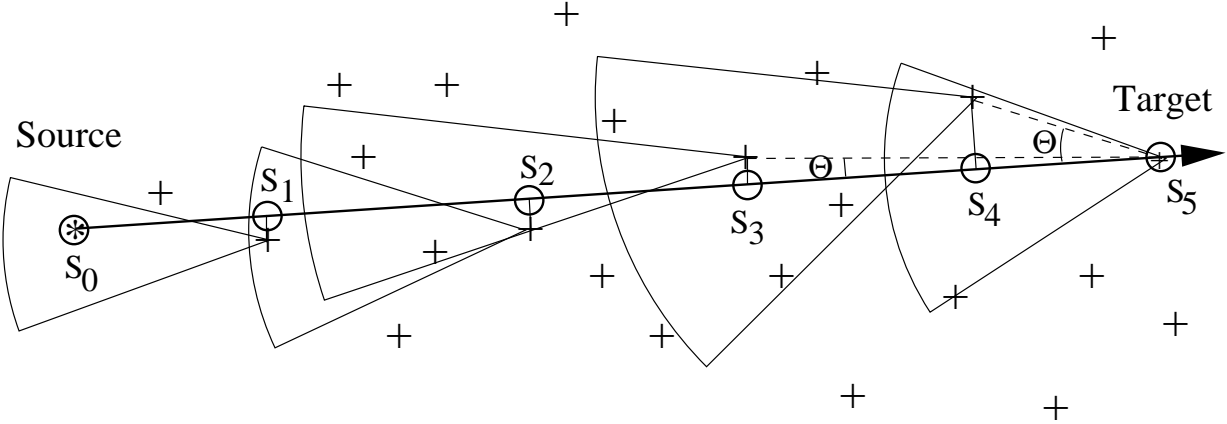}
  \caption{Illustration of the path finding procedure. Each plus sign
  represents a particle. The circle
  segments symbolize the radius of the volume filled with particles in the nearest
  neighbour list of the corresponding particle. The particle with the smallest angle $\Theta$
  between the line of sight from S0 to S5 and the line connecting them with the target are
  used for the determination of the evaluation points for the
  integration along the line of sight (small circles).} \label{fig:construct}
\end{figure*}

\subsection{Finding the evaluation points on the path towards the
  source} \label{sect:path}

First, we specify the position, the rate of ionizing photons $S_{\rm
tot}$ and $\bar{\sigma}$ (from Eq.~\ref{eq:crossect}) of the source.

For each particle~i we now proceed in the following way (see Fig.~\ref{fig:construct}):
Given the list of nearest neighbours of particle i, which has to be
determined anyway for the {\sc sph} formalism, we look for the particle j in
the list, closest to the line of sight defined by the smallest angle $\Theta$
between the line connecting the particles i and j and the line of
sight. We choose the angle between, not the distance from, the line of
sight, since we are interested in controlling the error in the
direction towards the source. This is not garanteed by the latter
criterion.

%%\begin{figure}
%%  \epsfig{file=crit.eps,width=8cm}
%%  \caption{Illustration of the tolerance angle criterion. Although
%%  d$_1$ is smaller than d$_2$, $\Theta_1$ is larger than
%%  $\Theta_2$. We are interested in controlling the error in direction,
%%  thus we choose $\Theta$ as our tolerance criterion.}
%%	\label{fig:criterion}
%%\end{figure}

We store this particle in a list and determine the evaluation point
S$_{\rm j}$ as the projected particle position on the line of sight. To
determine the next evaluation point S$_{\rm k}$ even closer to the
source we now repeat this method using the neighbor list of particle j
 and so forth until we reach the source.

%%First, the path from the particle~i to the source has to be
%%found. Helpful is here the list of nearest neighbours, which must be
%%determined for the {\sc sph} algorithm. We now look for the particle j in this
%%list, for which the LOS and the line connecting it and the particle~i
%%include the smallest angle $\theta$. We memorize the number of this
%%particle and the position of the crossing point of the LOS and the
%%vertical to it through particle j (see
%%Fig.~\ref{fig:construct}). We repeat this method with the particles
%%in the neighbour list of particle j etc, until at last we reach the source.

\subsection{Calculating the optical depth and ionization rate for the particles}

\subsubsection{Method A: {\sc sph} formalism method}

Now the path from the source to particle~i is known, and the integration of Eq.~(\ref{eq:LOS}) can be discretized by
using the evaluation points $S_{\rm i}$. The value for
$n_{\rm H}$ can be estimated by using the {\sc sph} smoothing formalism:
\begin{equation}\label{eq:densest}
  n_{\rm H}\left( {\bf r} \right) = \sum n_{\rm H,i} W \left({\bf r}-{\bf
  r}_{\rm i} \right),
\end{equation}
where the sum runs over the particle corresponding to the evaluation
point and its nearest neighbours. $W$ is the weight factor for each
neighboring particle provided by the smoothing kernel. We calculate the optical depth
along the line of sight by applying the Trapezian Formula, until we reach
particle~i:
\[
  \tau_{{\rm k}+1} = \tau_{\rm k} + \frac{1}{2} \bar{\sigma} \left( s_{{\rm k}+1} -
  s_{\rm k}\right) \left( n_{{\rm H},{\rm k}+1} + n_{{\rm H},{\rm k}} \right),
\]
with $s_{\rm k}$ being the position of the evaluation point on the line of sight.
Note that this treatment neglects the effects of scattering of the
ionizing photons by recombination or dust.

The distance between two successive evaluation points is smaller
or equal to the local smoothing length, which determines the largest
distance of the particles included in the nearest neighbour list as
well as the spatial resolution.
This guarantees that the line of sight integration of
Eq.~(\ref{eq:LOS}) is discretized into a reasonable amount of
substeps, consistent with the resolution given by the underlying
particle distribution.

The flux of ionizing photons at the position of particle~i into the
direction of photon propagation $\hat{\bf e}_s$ is then
given by:
\[
  {\bf J}_{\rm i} = J_0 \cdot
  \hat{\bf e}_s \cdot \exp \left( - \tau \left( s \right) \right).
\]

With the ionizing flux known at the particle positions, the
nabla operator in Eq.~(\ref{eq:ionrate}) can be calculated by the
{\sc sph} formalism. It is given for each particle i as the sum over its
neighbours:
\begin{equation}\label{eq:nabla}
  {\cal I}_{\rm i} = - \sum \frac{m_{\rm
  j}}{\rho_{\rm j}} {\bf J}_{\rm j} \cdot \nabla_{\rm i} W_{\rm j,i}.
\end{equation}
Now we are able to solve Eq.~\ref{eq:ioniztime}, which we write as:
\begin{equation} \label{eq:diffgl}
  \frac{{\rm d}x_{\rm i}}{{\rm d}t} = {\cal I}_{\rm i}-n_{\rm i}x_{\rm i}^2\alpha_{\rm B}.
\end{equation}
The time scale for the establishment of ionization equilibrium is
given by $1/(n \alpha_{\rm B})$, which is regularly much shorter than the
dynamical and gravitational timescales we are interested in. In order
to avoid small timesteps arising from the usage of explicit methods,
we use an implicit scheme.  
The first order discretization of Eq.~\ref{eq:diffgl} over a time interval $\Delta t$ is given by:
\begin{equation} \label{eq:firstorder}
x_{\rm i}^{{\rm n}+1} = x_{\rm i}^{\rm n}+\Delta t \cdot \left({\cal
I}_{\rm i}^{{\rm n}+1} - n_{\rm i}^{{\rm n}+1} x_{\rm i}^{{\rm n}+1} \alpha_{\rm B} \right),
\end{equation}
where the indices n and n$+1$ denote the values at the beginning and the
end of the actual timestep $\Delta t$, respectively. We already know
all the values on the right hand side from advancing the particles by
the {\sc sph} formalism, except the value for ${\cal I}_{\rm i}^{\rm n+1}$. Therefore a fully consistent implicit treatment is not feasible. We use the following guess for this value:
\begin{equation} \label{eq:implicit}
  {\cal I}_{\rm i}^{{\rm n}+1} = {\cal I}_{\rm i}^{\rm n} \cdot \frac{ 1-\exp \left(
  -n_{\rm i}^{{\rm n}+1} \bar{\sigma} a_{\rm i}^{{\rm n}+1} \left( 1-x_{\rm
  i}^{{\rm n}+1} \right) \right)
  }{ 1 - \exp \left( -n_{\rm i}^{{\rm n}+1} \bar{\sigma} a_{\rm i}^{{\rm
  n}+1} \left( 1-x_{\rm i}^{\rm n} \right) \right) }.
\end{equation}
In this equation, we assign an effective radius $a_{\rm i}$ to each
particle i proportional to the mean particle separation, given by
$a_{\rm i}=(M_{\rm i}/\rho_{\rm i})^{1/3}$. This is the estimate of
the size of a region with the particle mass $M_{\rm i}$ and density
$\rho_{\rm i}$. The factor with the exponentials on the right hand
side accounts for the effect of higher absorption and hence ionization
rate with decreasing ionization fraction. 

We must use the effective radius $a_{\rm i}$ in Eq.~\ref{eq:implicit}
instead of the smoothing length $h$, since the method works analogous
to implementations in grid codes. In contrast to the {\sc sph}
formalism, each particle now represents a volume of total mass $M_{\rm
i}$ and density $\rho_{\rm i}$, in which ionizing radiation enters on
one side and leaves on the opposite side. The size of this volume is
given by $a_{\rm i}$ as defined above. It is proportional to the
particle spacing.
 
In contrast, $h_i$ differs from the mean particle separation as it is
defined by the condition that there is a fixed number of neighbors
$N_{\rm neigh}$ of mass $M$
in the sphere with radius $2\, h_i$ and is thus given as
\[
  h_i = \left( \frac{3 N_{\rm neigh} M}{32 \pi \rho}
  \right)^{1/3}.
\]
It depends on $N_{\rm neigh}$ and can therefore not be used instead of
$a_{\rm i}$ in Eq.~\ref{eq:implicit}.

%%This shows that $h$ depends on the number of neighbours chosen and is
%%no measure for the size of a spherical volume filled with constant density
%%$\rho$ and containing mass $M_{\rm P}$. Instead $a_i$ as defined above
%%is the right measure we need in Eq.~\ref{eq:implicit}. For commonly
%%used values for $N_{\rm neigh} \simeq 50$, both quantities differ by the
%%factor
%%\[
%%  \frac{a_i}{h_i} = \left( \frac{32\,\pi}{3\, N_{\rm neigh}}
%% \right)^{1/3},
%%\]
%%which is of order unity. 

One consequence of the discretization of the ionization rate
equation is that the solution in ionized regions tends to oscillate
around the equilibrium value. 
In order to avoid small timesteps arising from this, we set the
ionization fraction $x$ of particles with an
$x > 0.95$ to the equilibrium value $x_{\rm E}$, which is defined by
${\rm d}n_{\rm e}/{\rm d}t=0$ in Eq.~\ref{eq:ioniztime}:
\[
  \frac{{\rm d}x}{{\rm d}t} = \frac{1}{n} \frac{{\rm d}n_{\rm e}}{{\rm
  d}t} = \bar{\sigma} (1-x_{\rm E}) J - n x_{\rm E}^2 \alpha_{\rm B} = 0.
\]
With $k=\bar{\sigma} J / (n \alpha_{\rm B})$ follows that
\[
  x_{\rm E} = \frac{1}{2} \left[ \left( k^2+4k \right)^{1/2}-k
  \right].
\]

This method works well in absolutely smooth, noise free particle distributions. However, if
one wishes to initially distribute the particles randomly in space,
one runs into problems. The sum in Eq.~\ref{eq:nabla} is very
sensitive to noisy particle distributions. Eventually the noise can be
so high, that the error of the sum introduced by noise reaches the order of the sum
itself. The ionization rate then locally drops below zero for some
particles, which can only be avoided by smoothing
the ionization rate spatially over several smoothing lengths. The
result is poor resolution. We circumvent this problem in method B.
 
\subsubsection{Method B: grid based method}

In this case, a different method is used to discretize the calculation
of the optical depth. We determine the positions of the evaluation
points i along the line of sight as described in Sect.~\ref{sect:path}
and calculate the hydrogen density $n_{\rm H,i}$ at these positions
using Eq.~\ref{eq:densest}. The path is then divided into pieces with
length $\Delta s_{\rm i} = (s_{\rm i+1}-s_{\rm i-1})/2$, assuming a constant
hydrogen density $n_{\rm H,i}$ along each interval. The optical
depth for one piece can then be approximated by
\[
  \Delta \tau_{\rm i} = \bar{\sigma} n_{\rm H,i} \Delta s_{\rm i}.
\]
These contributions to the optical depth are summed up until we reach
the position located one effective radius $a_{\rm i}$ before the
position of particle k. A first order approximation for the
ionization rate is now given by
\[
  {\cal I}_{\rm k} = \frac{J_{0}}{2 a_{\rm k} n_{\rm
  k}} \exp \left(
  \tau_{\rm k-a} \right) \left( 1 - \exp \left(-\Delta \tau_{\rm k}
  \right) \right),
\]
where $\tau_{\rm k-a} = \sum \Delta \tau_{\rm i}$ denotes the optical
depth one effective radius before the particle's position and $\Delta
\tau_{\rm k} = 2 a_{\rm k} n_{\rm H,k} \bar{\sigma}$ the optical depth
across the particle.

With the ionization rate derived above we solve the ionization rate
equation as described for case A. One can
easily show that Eqs. (\ref{eq:firstorder}) and (\ref{eq:implicit})
now give the exact implicit first order discretization for
Eq.~(\ref{eq:diffgl}). The solution now approaches the equilibrium
value $x_{\rm E}$ in the ionized regions without the instabilities mentioned in
method A. It is not necessary to set $x$ artificially to $x_{\rm E}$.

Method A seems to be the more consistent method since it uses
the {\sc sph} formalism for the calculation of ${\cal I}$. This is the
reason why it is also discussed in this paper. Nevertheless we prefer
method B due to its robustness against noisy particle distributions and higher
consistency concerning the integration scheme and have applied it to a
couple of test cases. 

\subsection{Computational effort}

\begin{figure}[h]
   \epsfig{file=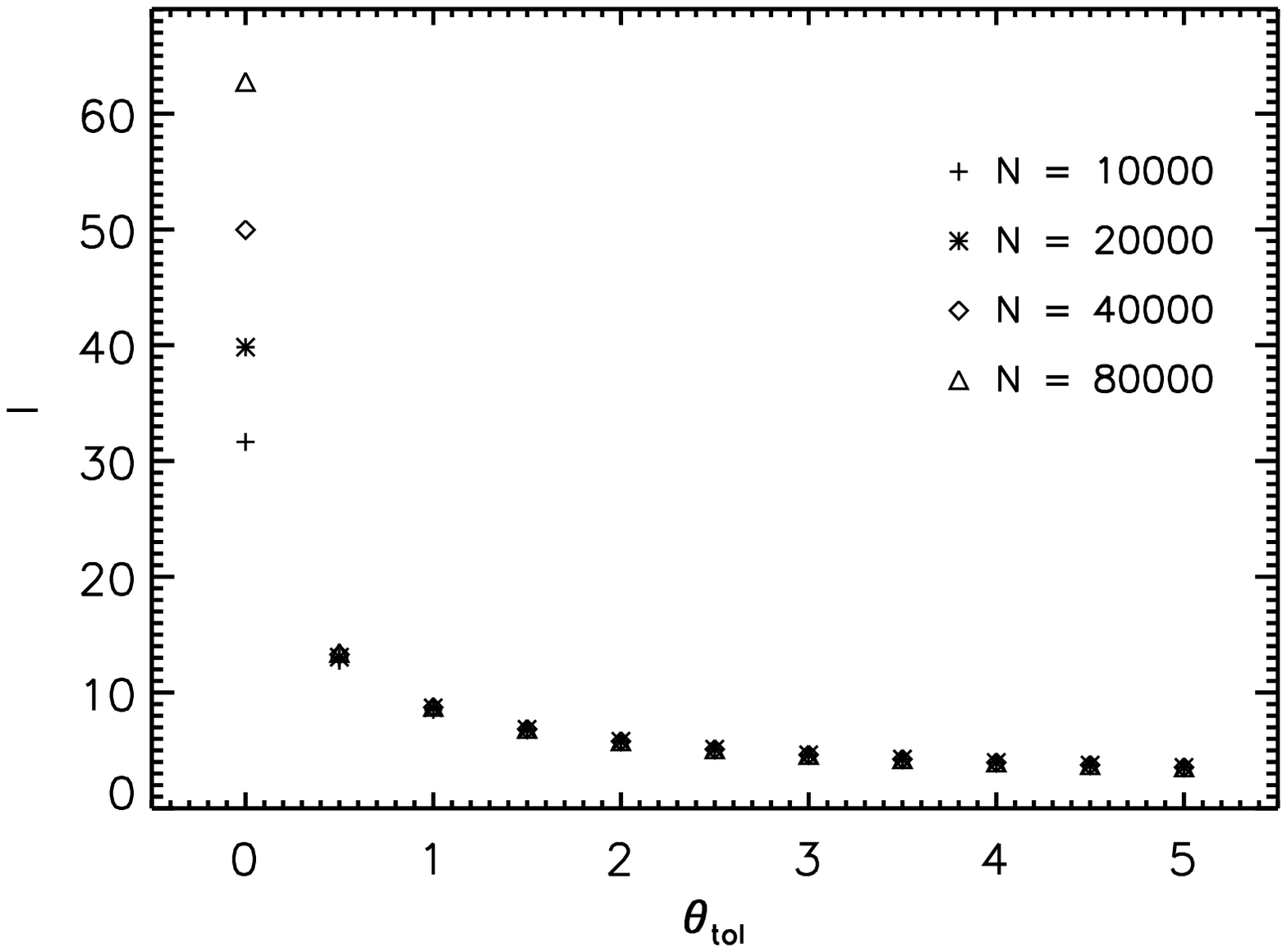,width=8.2cm}
   \epsfig{file=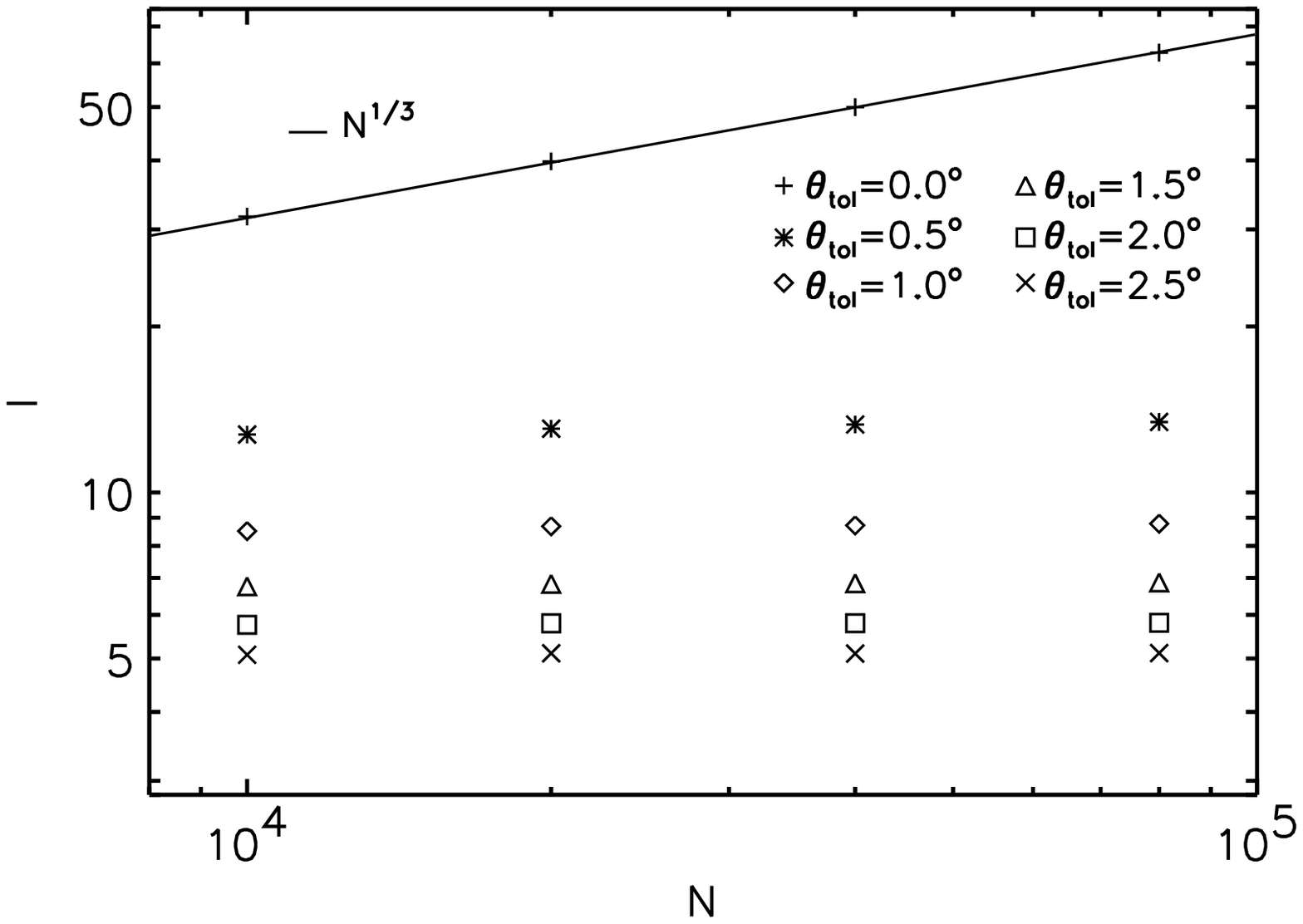,width=8.2cm}
  \caption{Mean number of evaluation steps $I$ per path depending on
   tolerance angle $\Theta_{\rm tol}$ and number of particles
   $N$. Upper panel: $I$ depending on $\Theta_{\rm tol}$ for different
   $N$. Note how $I$ drops with increasing $\Theta_{\rm tol}$. Lower
   panel: log-log-plot 
   of $I$ depending on $N$ for different $\Theta_{\rm tol}$ with
   the scaling law $I \propto N^{1/3}$ overplotted as a solid
   line. For $\Theta_{\rm tol} > 0^\circ$, $I$ becomes 
   independent of $N$ for large~$N$.}
  \label{fig:angdims}
\end{figure}

If the procedure explained above is used, the computational effort for
the line of sight integration scales approximately as $N^{4/3}$, since the
integration has to be done for each of the $N$ particles, and the
average number of evaluation points on each line of sight scales as
$N^{1/3}$. 

We can reduce the exponent from $4/3$ to 1 by introducing a
`tolerance angle' $\Theta_{\rm tol}$. Suppose we determine the particles along the line of sight as expalined . As soon as
$\Theta$ for a particle j along the line of sight towards the source is smaller
than $\Theta_{\rm tol}$ we stop our search here. The optical
depth of this particle $\tau_j$ is then used as an estimate of the
optical depth along the remaining part of the line of sight from the
source to S$_{\rm j}$. Thus no integration is needed for this part of
the path. One only has to make sure that $\tau_j$ is
already known, i.e. that the line of sight integration for particle j
has been performed earlier. In this case, the average number of
evaluation points $I$ per line of sight only depends on $\Theta_{\rm
tol}$ for large $N$. As shown in Fig.~\ref{fig:angdims}, $I$ becomes
constant for large $N$ and decreases with increasing $\Theta_{\rm
tol}$. As soon as $I$ becomes independent of $N$ the total
computational effort for all lines of sight together scales $\propto N$.

%%This is illustrated in Fig.~\ref{fig:angdims}. The upper panel shows
%%that, using the "tolerance angle", the mean number of evaluation steps
%%$N_{\rm mean}$ depends only on $\Theta_{\rm tol}$ and is decreasing with
%%increasing $\Theta_{\rm tol}$. As long as the mean number of
%%integration steps along the {\it whole} lines of sight is less or of
%%the same order as $N_{\rm mean}(\Theta_{\rm tol})$, 
%%tol})$ The lower panel shows that with increasing
%%particle numbers this mean value is approached faster for increasing
%%$\Theta_{\rm tol}$.

\begin{figure}
   \epsfig{file=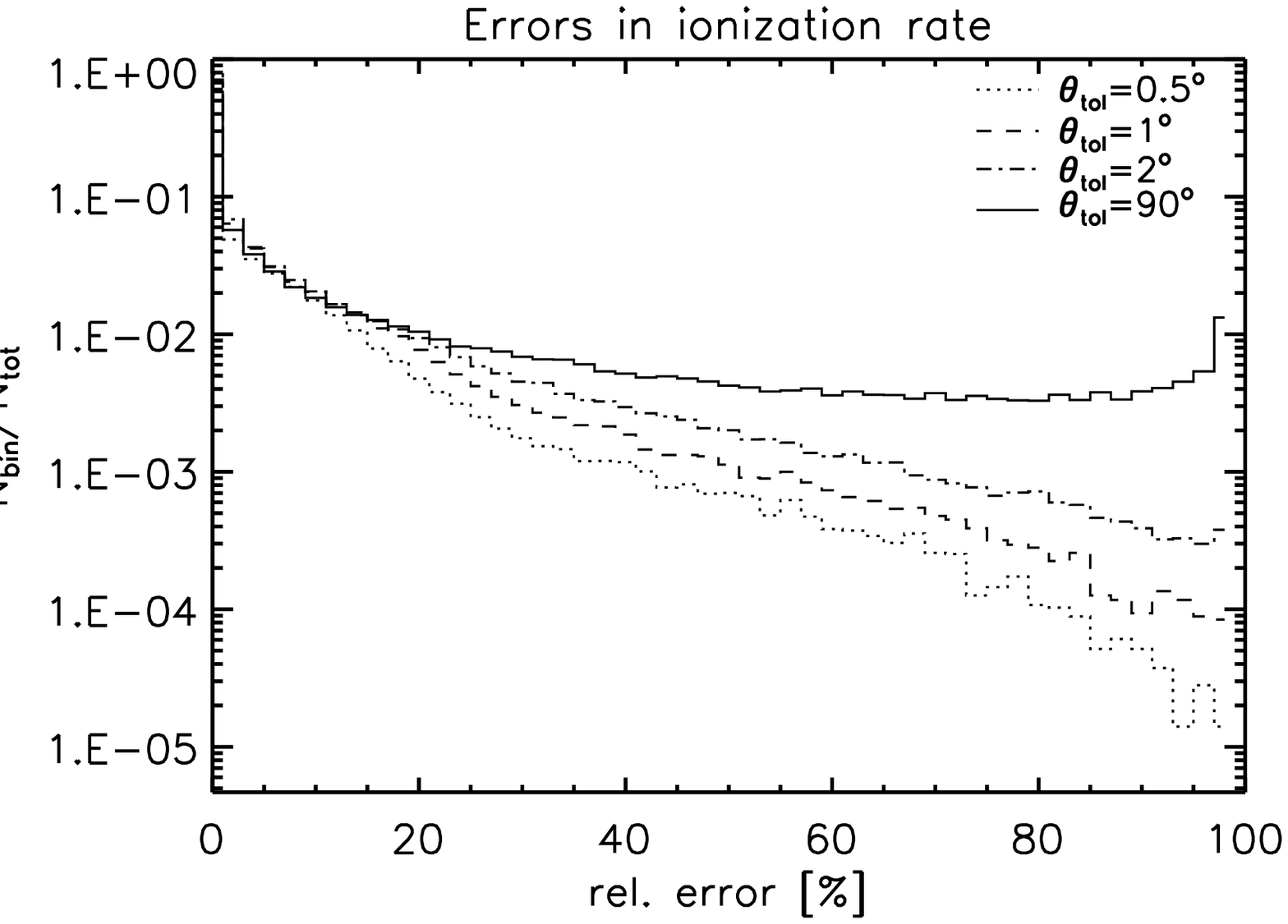,width=8.2cm}
   \epsfig{file=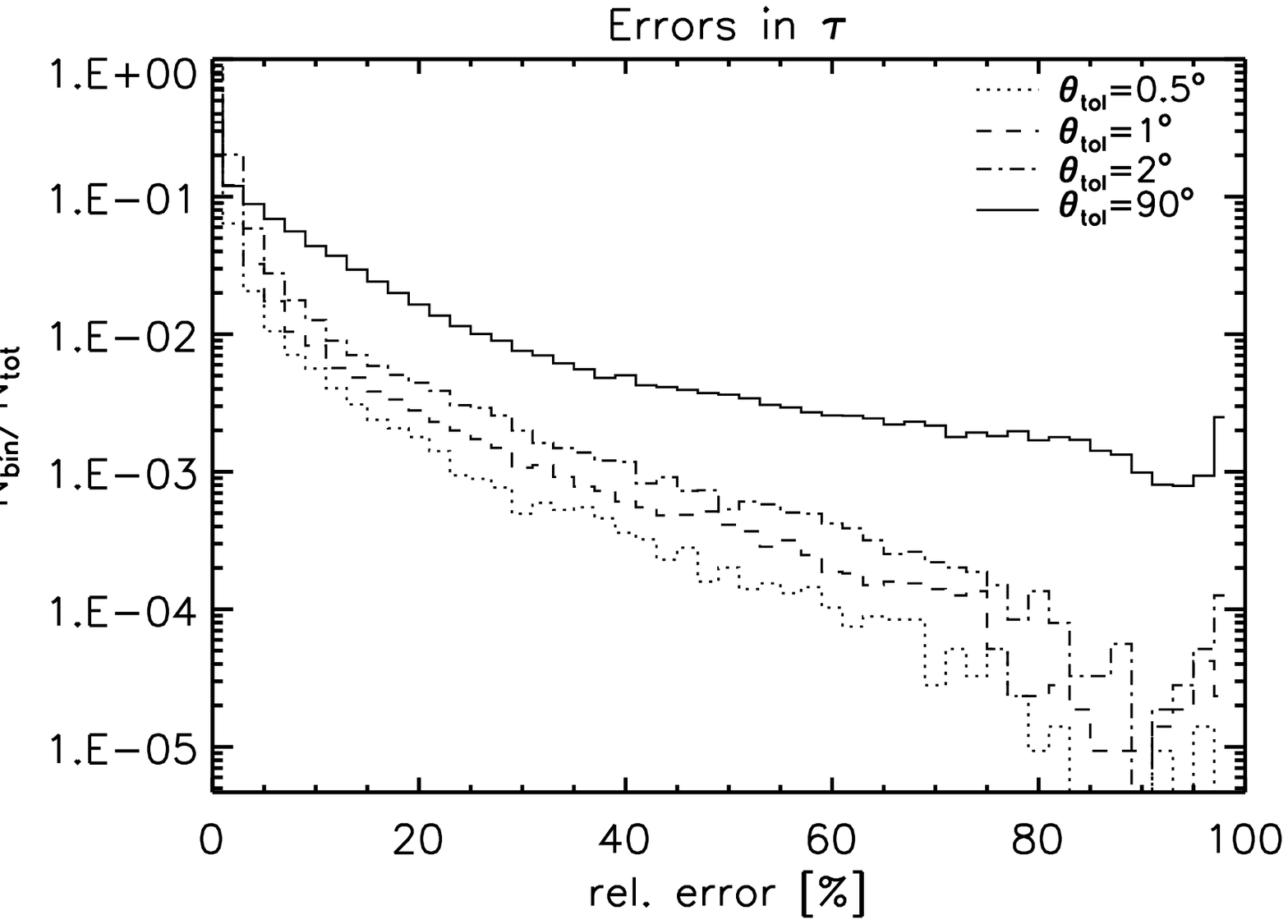,width=8.2cm}
  \caption{Histograms of the relative errors in $\tau$ and ${\cal I}$
  for different $\Theta_{\rm tol}$ in a three-dimensional test
  case.}
  \label{fig:raterrs}
\end{figure}

We demonstrate the effects of using the tolerance angle on the
accuracy of the ionization rate calculation in
Fig.~\ref{fig:raterrs}. Histograms are plotted for the errors in ${\cal I}$
and $\tau$ for calculations with
$\Theta_{\rm tol}=0.5^\circ$, $1^\circ$, $2^\circ$ and $90^\circ$
compared to $\Theta_{\rm tol}=0^\circ$. As the particle distribution we chose
the evolved state of a numerical simulation which studies the compression
and collapse of a dense clump within the UV field of an OB association
using 200\,000 particles. The results of this calculation will be
presented elsewhere \cite{kebu99}. Note that $\Theta_{\rm tol}=90^\circ$
represents the worst case, since the tolerance angle criterion now is
fulfilled for every particle with minimal $\Theta$ per search through the
nearest neighbour list. 

The particles which are most affected by the tolerance angle criterion lie
next to
the borders of shadows cast by optically thick regions, since here the path
for the integration along the line of sight may be bent through the optically
thick region, thus decreasing the ionizing flux artificially. In the
opposite case, the path may lead around the opaque region, increasing the ionizing flux
at the position of a particle in the shadow. These extreme cases lead to the
tail in the error histograms in Fig.~\ref{fig:raterrs}. Applying the
tolerance angle criterion thus numerically blurs shadows.        

The mean errors in $\tau$ are $1.3$ per cent for $\Theta_{\rm
tol}=0.5^\circ$, $2.2$ per cent for
$\Theta_{\rm tol}=1.0^\circ$, $3.4$ per cent for $\Theta_{\rm tol}=2.0^\circ$
and $11.2$ per cent for $\Theta_{\rm tol}=90^\circ$. The correspnding
mean errors in ${\cal I}$ are $2.8$, $4.1$, $5.7$ and $13.3$ per
cent, respectively. For the remaining test cases presented in this paper the choice
of $\Theta_{\rm tol}$ has no effect, since they deal with
one-dimensional problems, in which the optical depth is only a
function of distance from the source. Applying the tolerance angle
criterion only shifts the evaluation points from the lines of sight
in directions perpendicular to these, along which there is no change in
the optical depth. Indeed even the choice $\Theta_{\rm tol} = 90^\circ$
gives the same results in the one-dimensional test cases as for
$\Theta_{\rm tol}=0^\circ$. Thus the errors introduced by the angle
criterion must be checked with problems in which this symmetry is
broken and shadows are present, as the one mentioned above.

\subsection{Smoothing the ionization front}

For reasons of noise reduction we smooth the ionization front,
which is not resolvable by the {\sc sph} representation, over a distance of
the order of one local smoothing length. Nature provides a simple way for doing
this. The width of the ionization zone is of the order of one photon mean
free path length,
\begin{equation}\label{eq:width}
  d = (\bar{\sigma} \cdot n_{\rm H}) ^{-1},
\end{equation}
where $\bar{\sigma}$ is the net absorption cross section for ionizing
photons as defined in Eq.~\ref{eq:crossect}.

Since we cannot resolve the ionization region anyway, we are free to adjust $\sigma$ in a way that the width of the ionization
region given by Eq.~\ref{eq:width} is equal to a constant factor
$C \leq 1$ times
the local smoothing length $h$, but never larger than the value
$\bar{\sigma}$ given by Eq.~\ref{eq:crossect}:
\[
  \sigma = {\rm min}\left[\bar{\sigma},\left(n_{\rm H} \cdot C
  \cdot h\right)^{-1}\right].
\]
Test calculations have shown that a good value is $C=0.1$. It has
proven to sufficiently reduce numerical noise introduced into the
ionization structure by noise in the particle distribution and at the
same time to keep the resolution of ionization fronts better than the
resolution of the {\sc sph} formalism in order not to worsen the
overall resolution. Note that, when ``smoothing'' the ionization front over
$0.1$ times the smoothing length, the noise reducing effect is not
caused by the spatial smoothing, since it is ten times smaller than the {\sc
sph} smoothing. It rather results from a larger number of time steps
needed to ionize a particle in the front from an ionization fraction
of $x=0$ to $x \simeq 1$. This gives the neighbouring particles the
opportunity to react to the changed state in a smoother way.
 
\subsection{Heating effect}

We assume that heating and cooling effects lead to an equilibrium
temperature of 10\,000 K in the ionized gas penetrated by ionizing
radiation. The cross sections for elastic
electron--electron and electron--proton scattering are of the order
$10^{-13} {\rm cm}^2$. Together with a mean velocity of the electrons of
the order of $600 \rm{\,km\, s}^{-1}$ the thermalization timescale for the
energies of the ejected electrons is far less than a year for
densities of 1 particle cm$^{-3}$, which is many orders of magnitude
smaller than the dynamical timescale.  Thermalization thus occurs
quasi instantaneously. This process runs even more rapidly
for higher densities. Thus we are allowed to treat the gas behind the
ionization front as thermalized. We set the internal energy to:
\[
  e = x \cdot e_{10000} + (1-x) \cdot e_{\rm cold},
\]
with $e_{10000}$ being the internal energy corresponding to a
temperature of 10\,000 K for ionized hydrogen, and $e_{\rm cold}$ to the internal energy
for the 10 K cold, neutral gas. Note that this
method does not properly treat recombination zones,
since in this case one needs the correct inclusion of the heating and cooling
processes in order to achieve the correct gas temperatures, sound
velocities and
pressures. Also, the equilibrium temperature in H{\sc ii}-regions can
vary by 20 per cent from this value. These deviations can also only be taken
into account by proper treatment of heating and cooling.

\section{Tests of the numerical treatment}

Although being of one-dimensional nature, the following test
problems were performed fully in three dimensions.

\subsection{Test 1: Ionization of a slab with constant density}

With this problem we test the implementation of the time-dependent
ionization rate equation by ionizing a slab of H{\sc i} gas of
constant density $n$ with ionizing radiation falling perpendicular onto one of the boundary
surfaces. With hydrodynamics switched off, we let the ionization front
traverse the slab with a constant velocity $v_{\rm f}$.
To achieve this, we have to vary the infalling photon flux with
time. It is given by
\[
  J(t) = J_{\rm f} + J_{\rm t} = n v_{\rm f} + n^2 \alpha_{\rm B} v_{\rm f} t,
\]
where the first term on the right hand side is the flux which
provides the photons being absorbed in the ionization front.
The second, time-dependent term equals the loss of photons on
their way through the slab until they reach the front.

For the initial setup we place a number $N$ of
particles randomly into a slab with length-to-height and
length-to-width ratios of~10. Subsequently we let the particle distribution relax by evolving
it isothermally within the slab, adding a damping term to the force
law. This is necessary to diminish the numerical noise which was
introduced by the random distribution. We now have an ensemble of
the particles which does not possess any privileged directions and
which represents a gas of constant density and temperature. We
use this distribution as our starting configuration. From now on we
keep the particles fixed in space and switch off hydrodynamics.

%%\begin{figure}
%% \epsfig{file=parms.eps,height=6cm}
%%  \caption{Ionized mass vs. time for test 1. Solid line: theoretical
%%  solution. Results of calculations: Plus signs: 2000
%%  particles, stars: 16000 particles, diamonds: 128000 particles.
%%  Mass in unit of
%%  total mass $M_{\rm slab}$ in the slab. Time in units of time needed for the
%%  ionization front to cross the
%%  slab of length $L_{\rm slab}$ with a propagation velocity $v_{\rm f}$.}
%%  \label{fig:test1}
%%\end{figure}

\begin{figure}
  \epsfig{file=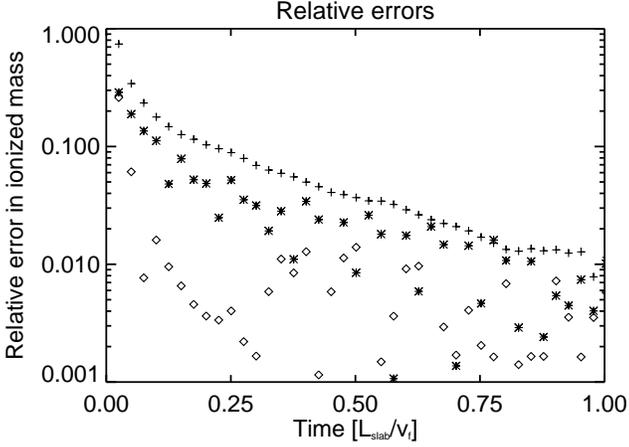,height=6cm}
  \caption{Relative error in ionized mass vs. time between
  calculations and theoretical result for test 1. Plus signs: 2000 particles,
  stars: 16000 particles, diamonds: 128000 particles. Time in units of
  time needed for the ionization front to cross the
  slab of length $L_{\rm slab}$ with a propagation velocity $v_{\rm f}$.}
  \label{fig:errors1}
\end{figure}
 
The test was performed for a total number of $N=2\,000, 16\,000$ and
$128\,000$ particles. Since the spatial
resolution for {\sc sph} calculations scales as $N^{-1/3}$ (with number of
neighbours $N_{\rm neigh}$ per particle fixed), this yields an increase
of linear resolution of a factor of two from one simulation to the
simulation with next higher resolution.
The results of these tests are shown in Fig.~\ref{fig:errors1}.

The mean relative errors between the theoretical result and the
calculations decrease linearly with increasing resolution, consistent
with our first order discretization of both the line of sight
integration and the time dependent ionization equation. The error also
decreases with time as the representation of the ionization front gets thinner and
thinner compared to the already ionized region. The spread in the errors for $N=16000$ and
$N=128000$ results from the fact that in these cases the numerical
solution oscillates around the theoretical solution, sometimes being
larger than the latter, sometimes smaller.

\subsection{Test 2: Ionization of a slab with density gradient}

We proceed as in test 1, with the difference that we choose a slab
with a constant density gradient in the direction of photon
propagation. We choose the time dependence of $J$ such that the
ionization front should travel through the gas with constant
$v_{\rm f}$. $J$ is given by:
\begin{eqnarray*}
J(t) = n_0 v_{\rm f} + \left(\alpha_{\rm B} n_0^2 + \frac{{\rm
d}n}{{\rm d} x} v_{\rm f} \right) v_{\rm f} t + \alpha_{\rm B} n_0 \frac{{\rm
d}n}{{\rm d} x} v_{\rm f}^2 t^2 + \\
\alpha_{\rm B} \left( \frac{{\rm
d}n}{{\rm d} x} \right)^2 v_{\rm f}^3 t^3,
\end{eqnarray*}
where $n_0$ denotes the density at the surface where the radiation
penetrates the slab and $\frac{{\rm d}n}{{\rm d} x}$ is the density gradient.

%%\begin{table}
%%\caption{Input parameters for test case 2.}
%%\label{tbl:test0}
%%\begin{tabular}{@{}lcl}
%%$n_0$ & : & 10 cm$^{-3}$ \\
%%$\frac{{\rm d} n}{{\rm d}x}$ & : & 495 cm$^{-3}$ pc$^{-1}$ \\
%%$v_{\rm f}$ & : & 1 pc Myr$^{-1}$ \\
%%$\alpha_{\rm B}$ & : & 2.59e-13 cm$^3$s$^{-1}$ \\
%%\end{tabular}
%%\end{table}

\begin{figure}
  \epsfig{file=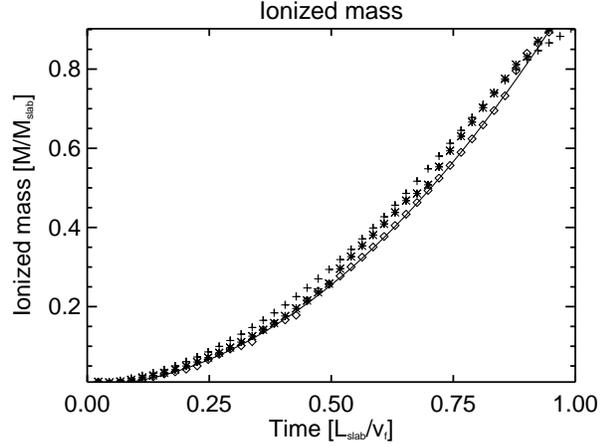,height=6cm}
  \caption{Ionized mass vs. time for test 2. Solid line: theoretical
  solution. Plus signs: N=2\,000. Stars: N=16\,000. Diamonds: N=128\,000.}
  \label{fig:test0}
\end{figure}

In Fig.~\ref{fig:test0} we plot the ionized mass for the theoretical
solution and the numerical simulations against time. The numerical
results converge against the theoretical solution with increasing
resolution. The deviations at $t>0.9$ are caused by the
ionization front reaching the rear boundary of the slab.

Note that the version of {\sc sphi} used in this paper is not able to follow
ionization fronts exactly which travel faster than one local smoothing
length per time step. This must be taken into account during the
timestep determination. In applications with fast ionization fronts
(typically R-type fronts in the early phases of the evolution of
H{\sc ii}-regions) this criterion can lead to very small timesteps and thus to
a high amount of CPU time needed. A version which circumvents this
problem is being developed.

%%\begin{figure}
%%  \epsfig{file=errorsgrad.eps,height=6cm}
%%  \caption{Errors in ionized mass vs. time for test 2. Solid line: theoretical
%%  solution. Plus signs: N=2,000. Stars: N=16,000. Diamonds: N=128,000.}
%%  \label{fig:errors0}
%%\end{figure}

\subsection{Test 3: Coupling of ionization and hydrodynamics}
\label{sect:test1}

\begin{figure}
  \epsfig{file=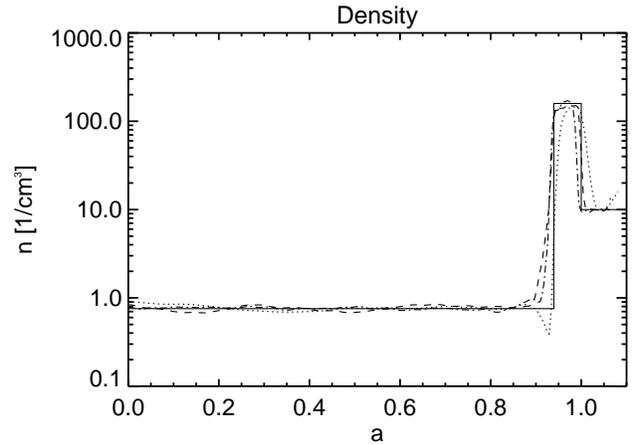,height=6cm}
  \caption{Density profile for test case 3 and different resolutions
  of the {\sc sphi} calculation. Unit of the x-axis normalized to the
  position of the shock front. Ionizing radiation infall from the
  left. A shock wave traveling to the right into the undisturbed
  medium with $n_0=10$ cm$^{-3}$ and $T_{\rm cold}=100$K sweeps up a dense shell of post-shock
  material, which is separated from the thin, hot, ionized material by
  an ionization front. Solid line: analytical result. Ratio of the
  thickness of the swept-up layer to the current local smoothing length for
  different resolutions: dotted 6,dashed 12, dash-dot 20. Corresponding
  times in code units: 0.34, 0.70, 1.0.}
  \label{fig:test2dens}	
\end{figure}

\begin{figure}
  \epsfig{file=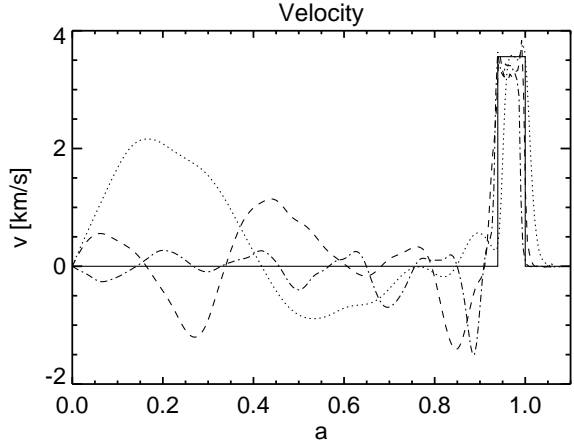,height=6cm}
  \caption{Velocity profile for test case 3 and different
  resolutions.  Unit of the x-axis normalized to the
  position of the shock front. Ionizing radiation infall from the
  left. Solid line: analytical result. Ratio of the
  thickness of the swept-up layer to the current local smoothing length for
  different resolutions: dotted 6,dashed 12, dash-dot
  20. Corresponding times in code units: 0.34, 0.70, 0.93, 1.0.} 
  \label{fig:test2vel}
\end{figure}

For this test we adopt the problem mentioned by Lefloch \&
Lazareff~\shortcite{lefloch1}. A box filled with atomic hydrogen of
particle density $n_0=10$ cm$^{-3}$ and temperature $T_{\rm cold}=100$\,K is exposed to ionizing
radiation, with the photon flux increasing from zero linearly with
time with a rate ${\rm d}\Phi/{\rm d}t=5.07 \cdot 10^{-2}$
cm$^{-2}$\,s$^{-2}$. There exists an analytical solution to this
problem, which is
self similar in the sense that physical values at position $x$ measured
in the direction of the photon flow at time $t$ are only functions of
$x/t$. This means: the structure is stretched with time. The
convergence of the code towards the correct solution with increasing resolution can be tested in
one calculation, since for all appearing structures the ratio between
structure sizes and smoothing lengths increases linearly with time.

The resulting structure is the following: an isothermal shock is
driven into the neutral medium, sweeping up a dense layer of
material. This is followed by an ionization front which leaves the
ionized material in quasi-static equilibrium (see Figs. \ref{fig:test2dens},\ref{fig:test2vel}). Using the parameter
$\Lambda=\alpha^{-1}({\rm d}\Phi/{\rm d}t)$, Lefloch \&
Lazareff~\shortcite{lefloch1} find the following analytical solution:
\begin{eqnarray*}
\Lambda=n_{\rm i}^2V_{\rm i} \\
n_{\rm i}=\left(\frac{n_0 \Lambda^2}{c_{\rm i}^2} \right)^\frac{1}{5} && V_{\rm i}=\left(\frac{\Lambda c_{\rm i}^4}{n_0^2} \right)^\frac{1}{5} \\
n_{\rm c}=n_0 \left(\frac{\Lambda}{n_{\rm i}^2 c_{\rm n}} \right)^2 && V_{\rm s}=c_{\rm n} \left(\frac{n_{\rm c}}{n_0} \right)^\frac{1}{2},
\end{eqnarray*}
where $n_{\rm i}$, $n_{\rm 0}$ and $n_{\rm c}$ denote the particle
densities of the ionized gas, the undisturbed neutral gas and the gas
in the compressed layer, respectively, and $V_{\rm i}$ and $V_{\rm s}$ the
velocities of the ionization front and the shock front, respectively.

We adopt $\alpha_{\rm B}=2.7 \cdot 10^{-13}$\,cm$^3$\,s$^{-1}$ from Lefloch \&
Lazareff~\shortcite{lefloch1} in order to directly compare the results of the
{\sc sphi} code to those of their grid-based method using a piecewise linear
scheme for the advection terms proposed by Van
Leer~\shortcite{VanLeer}. The resolution of 192 grid cells along
the slab of their calculations,
from which they derived their results, is comparable with the one used
in our high resolution case. We use the same method as described in
Sect.~\ref{sect:test1} to produce the initial conditions. No gas is
allowed to enter or leave the surface.

\begin{table}
\caption{Comparison of analytical and numerical results for test case 3.}
\label{tbl:result}
\begin{tabular}{@{}llll}
 & analytical & {\sc sphi} & Lefloch e.a. \shortcite{lefloch1}\\
\hline
$n_{\rm i}$ (cm$^{-3}$) & $0.756$ & $0.75\pm0.05$ & $0.748$ \\
$n_{\rm c}$ (cm$^{-3}$) & $1.59 \cdot 10^2$ & $(1.55\pm0.05) \cdot 10^2$ &
$1.69 \cdot 10^2$ \\
$V_{\rm s}$ (km s$^{-1}$) & $3.71$ & $3.67\pm0.05$ & $3.51$ \\
$V_{\rm i}$ (km s$^{-1}$) & $3.48$ & $3.43\pm0.05$ & $3.36$ \\ \hline
\end{tabular}
\end{table}

Table \ref{tbl:result} lists the result of this comparison. The {\sc sphi}
calculation slightly underestimates $V_{\rm s}$ and $V_{\rm i}$, as is
also observed for the grid code. The errors of order 5 per cent are
comparable to those achieved by Lefloch \&
Lazareff~\shortcite{lefloch1}.

\begin{figure}
  \epsfig{file=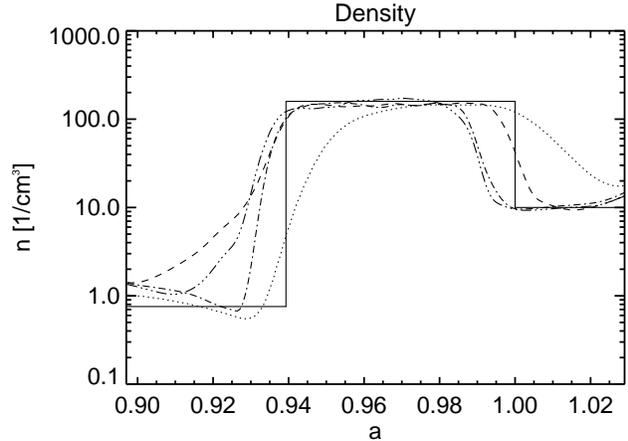,height=6cm}
  \caption{Density of the layer for test case 3. Unit of the
  x-axis normalized to the position of the shock front. Ionizing
  radiation infall from the left. Solid line: analytical
  result. Ratio of the thickness of the swept-up layer to the current local
  smoothing length for different resolutions: dotted 6,dashed 12,
  dash-dot 18, dash-dot-dot-dot 20. Corresponding times in code units:
  0.34, 0.70, 0.93, 1.0.}
  \label{fig:blow1}	
\end{figure}

\begin{figure}
  \epsfig{file=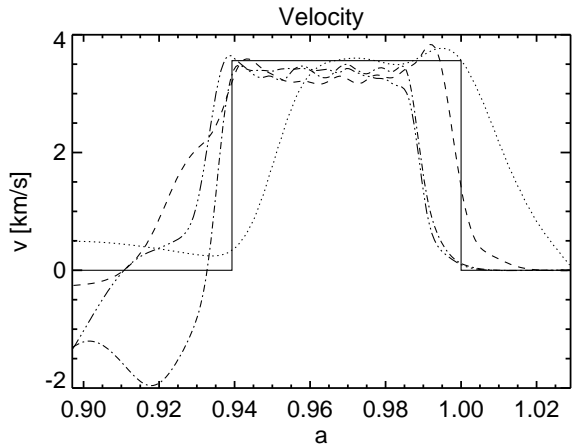,height=6cm}
  \caption{Velocity profile for test case 3. Unit of the
  x-axis normalized to the position of the shock front. Ionizing
  radiation infall from the left. Solid line: analytical
  result. Ratio of the thickness of the swept-up layer to the current local
  smoothing length for different resolutions: dotted 6,dashed 12,
  dash-dot 18, dash-dot-dot-dot 20. Corresponding times in code units:
  0.34, 0.70, 0.93, 1.0.} 
  \label{fig:blow3}
\end{figure}

In the early phases, i.e. low resolution, the
poor treatment of the ionization front leads to irregularities in the
ionized region and thus produces sound waves travelling back and forth
between the boundary to the left and the ionization front
(Figs.~\ref{fig:test2dens}, \ref{fig:test2vel}), which decrease in
power as time increases, i.e. at higher resolution.
With increasing resolution, i.e. increasing ratio of layer thickness
to smoothing length, the representation of the dense layer and the
shock front improves (Figs.~\ref{fig:blow1},
\ref{fig:blow3}).

\section{Summary}

The method presented in this paper allows the treatment of the
dynamical effects of ionizing radiation in {\sc sph}
calculations. Thus the study of astrophysical problems arising from
ionization, like the
impact of ionizing radiation from newly born stars onto the evolution of
their parental molecular clouds or the more consistent treatment of
heating by OB associations in galaxy dynamics calculations are now
feasible for the first time with {\sc sphi} in 3 dimensions. We demonstrate that the
code is able to treat time-dependent ionization, the related heating
effects and hydrodynamics correctly. Our first applications, 
detailed calculations of photoionization induced collapse in
molecular clouds and results obtained from them, will be presented in
a subsequent paper.

To allow the correct treatment of
recombination zones, one has to include the effects of time dependent
heating and cooling processes by ionization and recombination,
emission of forbidden lines and thermal radiation from dust. Another
important aspect which was neglected here is the effect of the diffuse
Lyman continuum recombination field. It can lead to the penetration of regions
shielded from the direct ionizing radiation by the ionization front,
which is e.g. seen in calculations of photoevaporating protostellar
disks \cite{yowe,richling}. An implementation
of these processes into our {\sc sphi} code is planned in the future.

\section*{Acknowledgments}
This work was supported by the Deutsche Forschungsgemeinschaft (DFG),
grant Bu 842/4. We'd also like to kindly thank Matthew
Bate and Ralf Klessen for the useful discussions concerning the
capabilities and implementation of the {\sc sph} method.

\end{document}